\documentstyle[sprocl,amsfonts,epsfig,rotating]{article}

\newcommand{\beq}{\begin{equation}}
\newcommand{\eeq}{\end{equation}  }

\newcommand{\dx}{\mbox{dx}}
\newcommand{\ds}{{\mbox{d}}\sigma}

\newcommand{\la}{\langle}
\newcommand{\ra}{\rangle}

\newcommand{\bec}{\begin{center}}
\newcommand{\eec}{\end{center}}

\def\nc{n_{\mathrm{c}}}
\def\nt{n_{\mathrm{t}}}
\def\sc{\sigma_{\mathrm{c}}}
\def\st{\sigma_{\mathrm{t}}}

\def\p+p{\pi^{\pm}\mbox{p}}
\def\K+p{\mbox{K}^{+}\mbox{p}} 
\def\m+p{\mu^{+}\mbox{p}}

\begin{document}

\title{GLUON DENSITY INSIDE THE  PROTON
FROM CURRENT-TARGET CORRELATIONS ? 
\footnote{Presented at XXVIII International Symposium on
Multiparticle Dynamics, Delphi, Greece, September 1998.} 
}

\author{S.~V.~CHEKANOV} 

\address{Argonne National Laboratory, 9700 S.Cass Avenue, \\
Argonne, IL 60439, \\ 
E-mail: chekanov@mail.desy.de}

\maketitle\abstracts{
The  possibility to determine the 
gluon density inside the proton  in deep inelastic ep
collisions using 
current-target multiplicity correlations is discussed. 
}  

\section{Introduction}

A significant fraction  of deep inelastic scattering (DIS) events 
contain  two jets in addition   
to the proton remnant.  They  are usually 
denoted  as (2+1) jet events. 
At small Bjorken $x$, the (2+1) jet events are predominately due to 
the Boson-Gluon Fusion (BGF) processes. Therefore, from a  study
of  (2+1) jet rates  one can learn about the 
gluon density inside the proton.  

The measurement of jet rates relies  upon various    
jet clustering algorithms (see a review \cite{clus}).   
The choice of  
jet algorithm  and  the associated resolution scale  
is, in some extent, arbitrary.  
The  jet algorithm  should
match  to a theoretical scheme used to calculate a cross section
and reflect as much as possible the parton  structure. 
However, there are 
some  factors  that  complicate the study  of the parton level
using  jet algorithms.   
Smearing effects and  ``misclustering'' are 
unwanted effects that are inherent to any algorithm. 
The  separation of perturbative  and non-perturbative  QCD
is also never perfect: 
The study of the hadronization contribution  to (2+1) jet rates 
is customarily based  on Monte Carlo (MC) models used to
compare  the parton and
hadron levels. The problem is that there is no uniquely defined
parton cascade in the MC simulations. A cut-off used to stop
the perturbative cascade is unnatural to QCD and can be different
for different MCs with different tunings.
This problem is compounded by the fact that the contribution from
the hadronization is also model dependent.     
Depending on the jet algorithm and Monte Carlo used, 
the hadronization corrections  to (2+1) jet cross-section
can vary from $15\%$  to $30\%$.

Besides BGF, the QCD Compton (QCDC) process  
gives  rise to  (2+1) jet events as well.
From  the jet algorithms themselves, this background,  
in principle,  cannot be isolated.  
For DIS, the remnant is
another  complication: The jet algorithms 
always suffer  from the problem of separation between the
spectator jet and the jets from hard QCD processes.  

Recently it was noticed that the BGF events can be studied  without  
involving  the jet algorithms \cite{chek}.  For this one 
can measure a liner interdependence between the
current- and target-region multiplicities  in  the Breit frame \cite{Br}. 
Since,  instead of clustering separate particles, 
the approach  involves the measurement of the particle  
multiplicities  in  large phase-space regions, 
one could expect  that  high-order QCD  
and hadronization effects are minimized.
Below we shall  discuss a few aspects of this method. 

\begin{figure}
\begin{center}
\begin{sideways}
\begin{sideways}
\begin{sideways}
\mbox{\epsfig{file=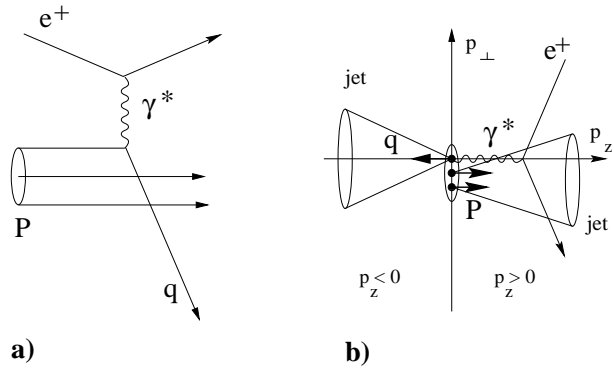, height=8.0cm}} 
\end{sideways}
\end{sideways}
\end{sideways}
\caption[brait1]
{a) Diagram for neutral current deep inelastic
scattering in the quark-parton model;
b) A schematic  representation of the Breit frame.
Particles with $p_z<0$ belong to the current region, 
particles with $p_z>0$ form the target region.}
\label{brait1}
\end{center}
\end{figure}

The DIS  processes can be characterized by the 
4-momentum transfer $Q^2=-q^2$ and the Bjorken
scaling variable $x=Q^2/(2 P\, q)$, where $P$ is the
4-momentum of the proton.
The fractional energy transfer $y$
is related to $x$ and $Q^2$
by $y\simeq Q^2/xs$, where $\sqrt{s}$ is the positron-proton
centre-of-mass energy.
For the quark-parton model in the Breit frame,
the incident quark carries $Q/2$ momentum
in the positive $z$-direction and the outgoing
struck quark carries the same momentum in the 
negative $z$-direction (Fig.~\ref{brait1}).
The phase space of the event can be divided into two
regions. All particles with negative $p_z^{\mathrm{Breit}}$ components
of momenta form the current region. In the quark-parton
model, all these particles are produced
from  hadronization  of
the struck quark. Particles with positive $p_z^{\mathrm{Breit}}$  are
assigned to the target region, which is associated with the
proton remnant.

\section{Current-Target Correlations. Analytical Estimates}
 
The correlation  between the current and target region
multiplicities  can be measured with the covariance  
\beq
\mathrm{cov}(\nc , \nt)= \la \nc\, \nt \ra -
\la \nc \ra \la \nt \ra,
\label{1}
\eeq
where $\nc$ ($\nt$) is  the number of particles in the
current (target) region. 
For the  first-order QCD effects leading  to    
(2+1) jet events,  the covariance receives a negative
contribution. At a fixed $Q^2$, the covariance can be 
written  as \cite{chek} 
\beq
\mathrm{cov}(\nc , \nt)   
\simeq  - A_1  R_1 ( x ) - A_2  R_2( x ),
\label{QCD}
\eeq
where $A_1$ and $A_2$ are positive,  $x$-independent constants. 
$R_1 (x)$ is the probability for  
back-to-back jet events with one jet in the current
and one in the target region and $R_2(x)$ is that
for an  event  without  jet activity in the current region
(both hard jets populate the target region). 
For small $Q^2$, both probabilities
are  mainly determined by the BGF events.  
The contribution from QCDC scattering is
relatively small because of the small fraction of such events
involved and since some QCDC events  have  
two hard jets in the current region, i.e. do not  produce   
such correlations.  

The parameters $A_1$ and $A_2$ determine the  average
value and width of the multiplicity distributions of the hard jets. 
Therefore, it is expected that
these quantities  are sensitive to  the 
higher-order QCD and hadronization
contributions.

\section{Hadronization}
\label{sec:mc}
 
We  illustrate some points concerning
the hadronization  using
the LEPTO 6.5 Monte Carlo model \cite{l65}. 
The model has been tuned as
described in \cite{l65t}. 
The hard process in LEPTO is described by a leading
order matrix element. 
The parton emission is based on  
the parton shower described by the  
Dokshitzer-Gribov-Lipatov-Altarelli-Parisi
evolution equation.
The JETSET Monte Carlo \cite{jetset} based on the
LUND string fragmentation model  is used to describe  
hadronization.      

To generate  DIS events, 
the energy of the positron and  that 
of the proton are   chosen to be  
27.5 GeV and  820 GeV, respectively. 
The  following cuts are used:
$10\> \mbox{GeV}^2 < Q^2 < 50\> \mbox{GeV}^2$,   
$E\ge 10\> \mbox{GeV}$, where $E$ is the energy of 
the scattered electron. 
No additional cuts for the track acceptance have been applied.  
For the given cuts, we investigate the correlations 
as a function of $x$.
For the given range of this variable, $\la Q^2 \ra$ varies  from
19.2 GeV$^2$  to 20.6 GeV$^2$. 
In total,  250k events  are  generated.

\begin{figure}
\begin{center}
\vspace{-3.0cm}
\mbox{\epsfig{file=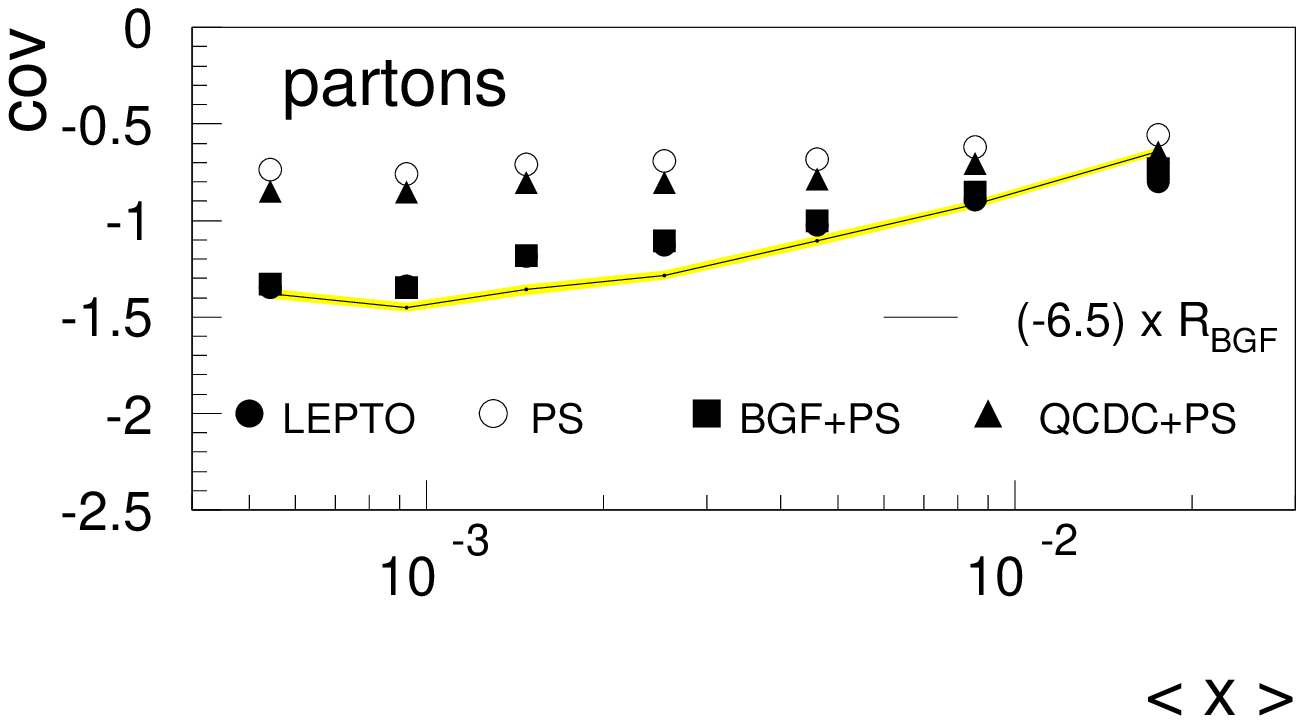, height=10.0cm}} 

\vspace{-2.0cm}
\mbox{\epsfig{file=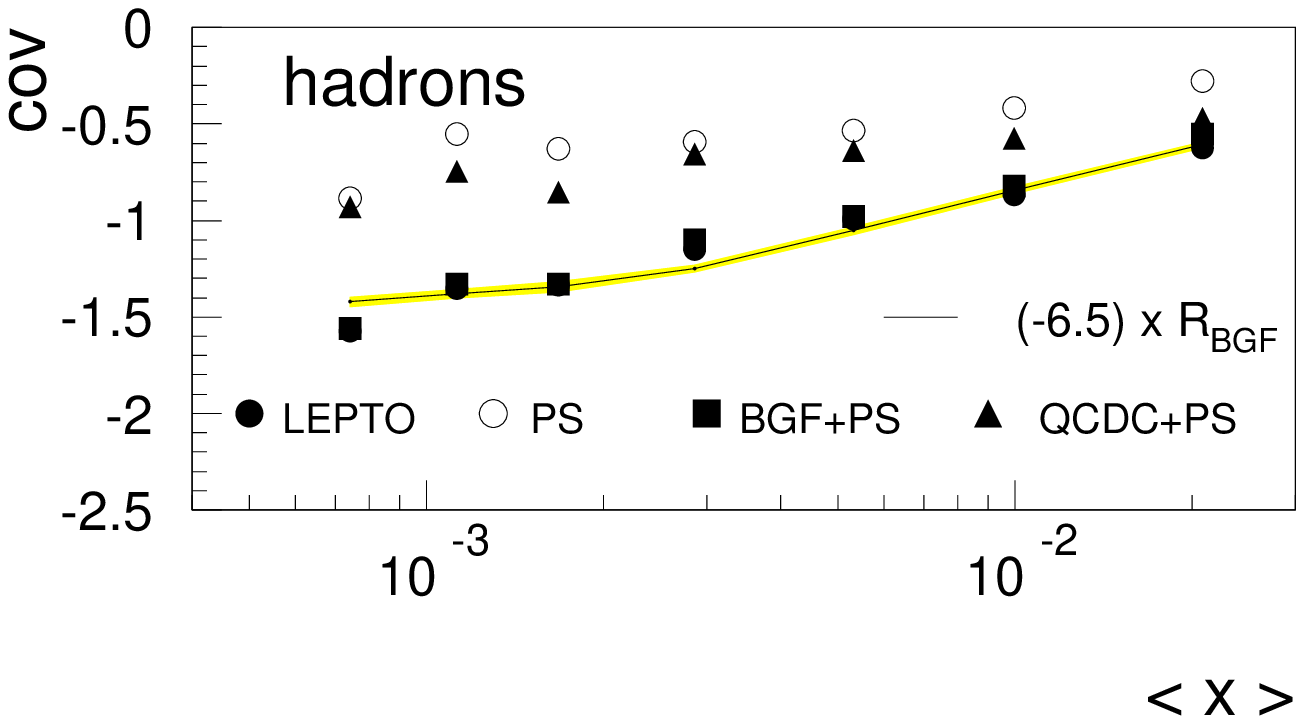, height=10.0cm}} 
\caption[fig1]
{Covariance for different bins
in $\la Q^2 \ra$ and $\la x \ra$ obtained from  the LEPTO 6.5
MC model on parton and hadron levels. The statistical errors
on the symbols are negligible. The shaded band on the line
shows  statistical uncertainties in 
the determination of the BGF rate.}
\label{fig1}
\end{center}
\end{figure}

The behavior of the covariance  
for partons  and hadrons  is   
shown in Fig.~\ref{fig1}.
Since the covariance is sensitive to the total multiplicity,
the QCD cut-off $Q_0$   used to stop
the parton cascade  is decreased from 
1 GeV to 0.68 GeV to obtain  about the same parton  multiplicity
in the current region as for hadrons.   
Open symbols show the parton shower model (PS) with  
hadronization (no first-order matrix elements).  
The parton shower with QCD Compton (QCDC+PS)
shows  about the same behavior 
since the number of the QCDC events is relatively small.  
The contribution from PS and QCDC is  nearly  independent of $x$. 
Note that without contributions from  QCD
processes, i.e. BGF, QCDC and PS
but including  the LUND
hadronization, the covariance is equal to 
zero \cite{chek} (not shown).

The correlations for PS and PS+QCDC are negative 
due to current-region particles with high $p_{\bot}$  
which have a large probability to migrate into the target region. 
This is in contrast to the remnants in the target region 
where no large $p_{\bot}$ is expected. From this consideration 
it is clear that the correlations for gluon radiation should
depend on $Q^2$, rather than on $x$. 
  
The covariance receives an additional  
negative contribution from  
BGF events (BGF+PS, closed squares in Fig.~\ref{fig1}).
The closed circles show LEPTO with all first-order QCD effects.
Since the results for the default LEPTO and LEPTO with
BGF+PS are very similar,  
we conclude  that the behavior of the current-target
correlations is dominated  by the BGF. 

The line shows the BGF rate $R_{\mathrm{BGF}}$ rescaled
using the constant -6.5. 
The rate is obtained from LEPTO by counting events
labeled as the BGF.     
The  correlations follow the BGF rate rather
well,  in spite  of the background from PS 
contributing to the absolute magnitude of the correlations
(see more examples in  \cite{chek}).   

No large difference  between the parton and hadron levels 
is observed. 
This illustrates  the fact that the LUND string model 
does not produce a strong effect on the current-target correlations.
Some effect, however, is seen for the covariance measured at  
the smallest $x$ value. It could be that this effect comes from
the strings connecting the partons from current and target regions.
Since the remnants have  high 
longitudinal momentum $\sim Q/2x$, opposite
to relatively small $x$-independent 
momenta  of the current-region partons, the
strings should carry particles away from the current    
to the target region, 
producing negative correlations  which could be seen at a sufficiently
small $x$.  From the MC study, however, it is seen that this effect 
is relatively small for the experimentally  accessible $x$ region.  

\begin{figure}
\begin{center}
\vspace{-3.0cm}
\mbox{\epsfig{file=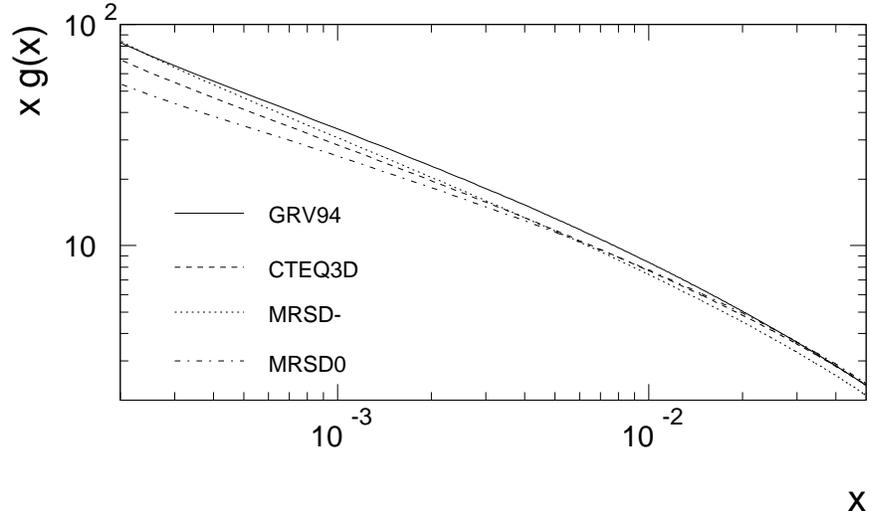, height=10.0cm}}
 
\vspace{-2.0cm}
\mbox{\epsfig{file=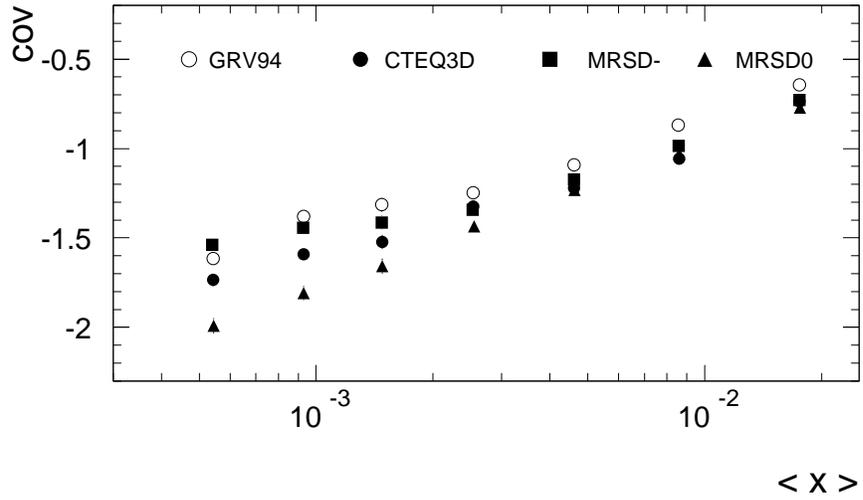, height=10.0cm}}
\caption[fig4]
{Different gluon densities and corresponding 
current-target correlations as a function of $\la x\ra$ 
for LEPTO with the default parameters. 
Note that in LEPTO 
an increase of the gluon density leads
to a {\em decrease} of the BGF rate. This trend drives the behavior
of the current-target correlations shown on the bottom  figure.}
\label{fig2}
\end{center}
\end{figure}

\section{Structure-Function  Study}

The MC results show that 
the current-target correlations are  approximately  
proportional to the rate of BGF events. 
The same conclusion follows 
from the analytical estimates showing a linear 
dependence of the correlations on  $R_1$ and $R_2$
probabilities determined by BGF rate at small $Q^2$. 
Therefore, one might  expect that
the correlations  are sensitive to the behavior of the 
parton density. Fig.~\ref{fig2}  shows   different
gluon  densities obtained from the PDFLIB \cite{PDFLIB} 
and the corresponding current-target
correlations. The sensitivity of the covariance 
to the input structure functions is apparent. 
Note that the  gluon densities with a steeper 
increase do not  lead to stronger correlations   
since the covariance is  determined 
by an interplay between 
the BGF cross-section and the total differential cross section,
\beq 
\mathrm{cov} (\nc , \nt)\mid_{Q^2=cons} \> 
\propto - \frac{\ds_{\mathrm{BGF}} /\dx}
{\ds /\dx}. 
\label{QCD2}
\eeq
Note that the behavior of the BGF 
rate for the different gluon
densities  has
the same trend (not shown) as the correlations. 
This trend is Monte-Carlo dependent:  
The BGF cross-section in
(\ref{QCD2}) is evaluated in LEPTO using cut-offs
to prevent divergences in the QCD matrix elements.

As we have discussed, the coefficient 
of proportionality in (\ref{QCD2}) is determined by the
structure of the multiplicity distributions of the outgoing quarks.
Therefore, it
absorbs non-BGF effects, high-order QCD corrections 
and hadronization effects. Note also 
that relation  (\ref{QCD2}) can be used to
determine the $x$-behavior of $\ds_{\mathrm{BGF}} /\dx$ from  
the measured $\mathrm{cov} (\nc , \nt)$ and the overall differential
cross section.  
 



\section{Discussion: Open Questions}

{\it Theoretical aspects.}
As for any  measurement of the differential (2+1) cross section,
it is important to understand the Monte Carlo dependence of the 
discussed results. 

A recent ZEUS study \cite{ZEUS} 
showed  that only  ARIADNE \cite{ARD}  
can quantitatively describe the current-target
correlations. Nevertheless, comparing different MC results, one could
see that all MCs show about same $x$-behavior. 
ARIADNE, however, shows
systematicly larger values of the covariance.  
This observation is  consistent with  the assumption that all
differences in the treatment of the parton showers 
and hadronization stage are absorbed  into the coefficient  
of the proportionality  in (\ref{QCD2}), rather than contribute
to the $x$-dependence of the correlations.     
  
From  a perturbative QCD point of view,
the higher-order effects are  not sufficiently understood.
For example, for  the next-to-leading order QCD calculations, the two-jet
cross section receives contribution from the 3-parton final state.
The hope is that, to  a large extend, the
behavior (but not the magnitude) of the correlations stay the same.  
For example, ``unresolvable'' partons   mainly  determine the
structure of the jets, rather than  their  locations
in the Breit frame.   

\vspace{0.5cm}
{\it Experimental aspects.} 
The most important experimental question is how to measure 
the current-target correlations having a detector with a small track
acceptance for the target region. 
The absolute value of the 
covariance is sensitive  to  the fraction of tracks measured
in the target  region. In this respect, there are two possibilities.    
Since we are interested in the $x$-behavior of BGF cross-section,
one could assume that a small track acceptance cannot affect such a behavior,
but rather the absolute normalization 
which is absorbed into a unknown coefficient of 
proportionality in (\ref{QCD2}). This question should be carefully
examined using Monte Carlo simulations.  
Another  approach  could be the use of the other characteristics
of the correlations which are less  sensitive to the mean value 
of the multiplicity distribution measured in the target region. 
For example, the coefficient of
the correlation $\sc^{-1} \st^{-1} \mathrm{cov}$, 
($\sc$ and $\st$ being the standard deviations of the multiplicity
in the current and target regions) could be useful,  
once the $x$-behavior is   understood. 

\vspace{-0.3cm}
\section*{Acknowledgments}

\vspace{-0.2cm}
I thank M.~Derrick, E.~De Wolf, D.~Krakauer, S.~Magill, 
J.~Repond for 
helpful discussions and corrections.

{}

\end{document}